# Highly Active Nanoperovskite Catalysts for Oxygen Evolution Reaction: Insights into Activity and Stability of $Ba_{0.5}Sr_{0.5}Co_{0.8}Fe_{0.2}O_{2+\delta}$ and $PrBaCo_2O_{5+\delta}$

Bae-Jung Kim, Xi Cheng , Daniel F. Abbott, Emiliana Fabbri, Francesco Bozza, Thomas Graule, Ivano E. Castelli, Luke Wiles, Nemanja Danilovic, Katherine E. Ayers, Nicola Marzari, Thomas J. Schmidt

**Abstract**

It is shown that producing $PrBaCo_2O_{5+\delta}$ and $Ba_{0.5}Sr_{0.5}Co_{0.8}Fe_{0.2}O_{2+\delta}$ nanoparticle by a scalable synthesis method leads to high mass activities for the oxygen evolution reaction (OER) with outstanding improvements by 10× and 50×, respectively, compared to those prepared via the state-of-the-art synthesis method. Here, detailed comparisons at both laboratory and industrial scales show that $Ba_{0.5}Sr_{0.5}Co_{0.8}Fe_{0.2}O_{2+\delta}$ appears to be the most active and stable perovskite catalyst under alkaline conditions, while $PrBaCo_2O_{5+\delta}$ reveals thermodynamic instability described by the density-functional theory based Pourbaix diagrams highlighting cation dissolution under OER conditions. *Operando* X-ray absorption spectroscopy is used in parallel to monitor electronic and structural changes of the catalysts during OER. The exceptional BSCF functional stability can be correlated to its thermodynamic meta-stability under OER conditions as highlighted by Pourbaix diagram analysis. BSCF is able to dynamically self-reconstruct its surface, leading to formation of Co-based oxy(hydroxide) layers while retaining its structural stability. Differently, PBCO demonstrates a high initial OER activity while it undergoes a degradation process considering its thermodynamic instability under OER conditions as anticipated by its Pourbaix diagram. Overall, this work demonstrates a synergetic approach of using both experimental and theoretical studies to understand the behavior of perovskite catalysts.

# 1 Introduction

Renewable energy sources show great potentials to alleviate the ongoing environmental threats from the non-renewable energy sources. However, the implementation of renewable energy technologies is challenged by the intermittent production of electricity, which therefore requires an efficient energy storage system to mediate generation and consumption of energy. In this respect, the growing needs to store larger amounts of energy have recently angled the spotlight towards water electrolysis technologies.**1** For an efficient water splitting process, an effective oxygen evolution reaction (OER) is required at the anode.**2** In recent developments, members of the perovskite oxide family have been recognized as highly active electrocatalysts capable to facilitate an efficient water oxidation reaction relieving the need of rare and expensive precious metals.[**2**, **3**] The perovskite oxide family has the general chemical formula –$ABO_3$– composed of rare-earth (e.g., Pr and La) or earth alkaline metal (e.g., Ba and Sr) in the A-site and 3d transition metals in the B-site (e.g., Co). Perovskite oxides offer a broad window of opportunities to explore different synergetic effects between various non-precious metals in a crystalline structure owing to their capability to accommodate cation substitutions in both the A- and B-sites with another cation element, resulting in $(A_xA'_{1-x})(B_yB'_{1-y})O_3$ compositions.[**3**] Thus, one can benefit from substitutions that yield modified intrinsic properties, such as band structures and thus affecting electrical, optical, and magnetic properties, and ultimately tailoring the catalytic activity.**4** This compositional modification of perovskites can take place in either an ordered or a random arrangement within a unit cell. To date, $PrBaCo_2O_{5+\delta}$ (PBCO)[**3**, **5**] and $Ba_{0.5}Sr_{0.5}Co_{0.8}Fe_{0.2}O_{2+\delta}$ (BSCF)[**2**, **3**, **5**, **6**] have been revealed as two of the most active perovskite catalysts for OER, each representing a different structure: a layered order (also known as double perovskite) and a random arrangement (single perovskite), respectively. In particular, these cobalt-based perovskites have shown remarkable affinity towards oxygen evolution owing to the cobalt's unique ability to accommodate different oxidation states.**7**

The importance of fundamental understanding of reaction mechanisms cannot be stressed enough as it is the key for designing an optimal perovskite OER catalyst. In early works, Bockris proposed a variety of plausible OER mechanisms assuming that there is one rate determining step for each proposed mechanism based on experimentally determined Tafel slopes.**8** In recent years, quantum-mechanical computational methods, such as density-functional theory (DFT), have become a precious tool for thermochemical analyses to unravel catalytic mechanisms of oxygen evolution.[**2**, **9**] This approach derives its accuracy by studying trends across different catalysts and on the discovery that kinetic barriers scale with thermodynamic barriers, which allows the understanding of kinetics from a thermodynamic basis.[**2**, **10**] In this regards, detailed intermediates and transition states are still required in order to gain a precise understanding of catalyst's OER.[**10**]

As to acquire a deeper understanding of the complex mechanism behind OER of these active perovskite catalysts, o*perando* X-ray absorption spectroscopy (XAS) offers an effective way to monitor changes in catalysts during operation.**11** Our recent study takes advantage of this technique to investigate the local electronic and geometric structural changes of BSCF during OER.[**3**] In this study, the unique growth of a *self-assembled* B-site metal oxy(hydroxide) layer (i.e., Co/Fe–OOH) at the surface triggered by lattice oxygen evolution reaction (LOER) of the perovskite has been identified as the active species for OER on BSCF—a cubic perovskite.**3**, **12** Recent studies**13** highlight the importance of the LOER process for an efficient oxygen evolution. Given the BSCF's unique structural flexibility, the dissolved cations during the LOER process are dynamically reconstructed as Co/Fe oxy(hydroxide) active layer from which oxygen is partially evolved; simultaneously oxygen is also evolved from the electrolyte.[**3**, **12**] Alternatively, PBCO is structurally distinct from BSCF in its ordering of cations such that the lanthanide metal and the alkali-earth metal planes are alternating while only Co is the only transition metal that occupies the B-site. PBCO has been found to consist of a lower oxygen binding strength leading to a faster oxygen surface exchange rate and is able to accommodate swift ion migration.**14** Given its difference in the ordering of cations than BSCF, PBCO may reveal a different mechanistic pathway during OER which is yet to be elucidated.

In the last decade, number of descriptors[**3**, **5**, **13**, **15**] have been reported appraising the high OER activity of PBCO compared to other single perovskite catalysts. Particularly, PBCO's high activity towards OER was attributed to its O *p*-band center being adequately close to the Fermi level, which would be able to activate the LOER leading to a different reaction pathway than the interaction between adsorbents at surface meatal sites.[**3**, **16**] Also in these previous studies, PBCO has been considered to be a stable catalyst without cation dissolution and not undergoing amorphization during the oxygen evolution process in alkaline electrolyte, while there has been indications of oxygen participating from the oxide (i.e., LOER). However, the assessment of intrinsic properties of an electrocatalyst may not truly be revealed in a bulk scale considering that OER is an interfacial reaction that takes place at the surface of the catalyst material. Conventionally, sol–gel (SG) synthesis has been acknowledged as the state-of-the-art method to prepare metal oxides with high phase purity. Nevertheless, the resulting particle size from this synthesis method generally scales from few to several micrometers after the subsequent thermal treatment and calcination. From a practical application standpoint, it is tremendously advantageous to have nano-sized catalyst particles to yield high active surface area, for which the catalyst mass activity is maximized. Therefore, this work takes benefit of flame spray (FS) synthesis**17** to produce nano-sized catalysts with remarkably high surface areas yielding not only high OER mass activities but also enabling better extraction of catalytic properties during electrochemical reactions.

Overall, this study aims to provide a thermodynamic perspective in understanding the activities and functional stabilities of two among the most active perovskite catalysts for OER—PBCO and BSCF—combining with *operando* XAS to apprehend and compare their dynamic local electronic geometric structures under OER conditions. DFT computed Pourbaix diagrams are constructed to complement experimentally observed performances of the catalysts based on their thermodynamic nature. Furthermore, the selected perovskites are adopted in an operational alkaline exchange membrane water electrolyzer (AEMWE) cell as anode materials to exploit their practical performance in the technical application. Overall, we provide mechanistic overview behind the OER processes of BSCF and PBCO catalysts in alkaline environment. Moreover, we provide insights into employing BSCF and PBCO as the anode catalyst for a long-term functionality.

# 2 Results and Discussion

Nanoparticulates of PBCO and BSCF are prepared via FS synthesis. Each of the constituent metal precursors are dissolved in a combustible solution and injected into the flame at a high temperature (possible up to ≈3000 °C) and the resulting precipitates are collected. The phases of FS synthesized perovskite oxides are identified using X-ray diffraction (XRD), and compared to those of the same composition prepared via SG synthesis (Figure S1, Supporting Information). First referring to Figure S1a in the Supporting Information, BSCF produced from both synthesis methods reveal distinct peaks that are well indexed to those of the conventional BSCF [ICSD Collection Code: 162 269], yet the XRD peaks of BSCF prepared via FS synthesis (BSCF-FS) reveal a broader width than those of BSCF prepared via SG method (BSCF-SG). Note that some of the non-indexed XRD peaks indicate minor quantities of secondary phases, mostly carbonate species (e.g., $BaCO_3$). Also, these peaks of BSCF-FS are revealed at a lower 2θ angle than those of BSCF-SG. The earlier appearance of XRD peaks may be attributed to an increase in lattice parameter, which is typically ascribed to the presence of a reduced Co oxidation state in BSCF-FS.[2] XRD patterns of PBCO prepared via FS and SG (Figure S1b, Supporting Information) shows similar qualitative difference as BSCF. Given that FS products display broader XRD peaks owing to smaller crystalline sizes, XRD peaks in PBCO-FS displays convolution of some peaks located closely to another. Consistent with recent study,[3] and the FS synthesis method appears to promote accommodation of reduced Co (i.e., Co(II) rather than Co(III)) within the perovskite structure for a superior OER activity in alkaline electrolytes.

The physical traits of the perovskites prepared via FS and SG are compared using high-resolution transmission electron microscopy (HR-TEM). **Figure 1**a,b displays HR-TEM images of BSCF-SG and PBCO-SG, respectively. Although both perovskites prepared via SG method reveal pristine fringes ascribable to each of its specific crystalline structure, it is important to draw the attention to their large particle sizes (refer to insets of Figure 1a,b), with particle sizes that are in the average range of 3–5 µm. On the contrary, TEM images of BSCF-FS and PBCO-FS (Figure 1c,d, respectively) reveal nanosized particles (5–20 nm) as primary particulates retaining clear fringes (Figure 1c,d, insets). Consequently, these nanoscaled particles obtained from FS method exhibit significantly higher Brunauer–Emmett–Teller (BET) surface areas (refer to Table S1, Supporting Information), resulting in outstanding mass activities towards the OER. Figure 1e,f each compares OER activities in 0.1 M KOH in Tafel plots for the catalysts prepared by the two synthesis methods. The FS synthesized BSCF and PBCO clearly reveal a remarkable increase in current densities at 1.55 $V_{RHE}$ by factors of 57 and 10, respectively. Note, both perovskite oxides shows different Tafel slopes between SG and FS synthesis, which are related to their physico-chemical differences of the catalysts produced by each method (refer to Figure S4, Supporting Information).[3] Also, notice that their improvements of OER activities are not proportional to their increases of surface areas. This suggests that the increase of surface area owing to having nanosized particles is not the only contribution to the enhanced mass activity. In this respect, previous studies[2, 5, 18] have shown a strong relationship between OER activity of cobalt-based perovskites and their cobalt oxidation states; particularly, a higher OER activity was achieved by BSCF when its Co was in a more reduced oxidation state as an effect of carbon accompaniment. Also confirmed by our recent study,[3] BSCF with a reduced Co oxidation state accommodating a large number of oxygen vacancies without the aid of carbon also revealed a high-mass activity compared to stoichiometric BSCF. Similarly, in the case of PBCO, the increase of the current density could be partially attributed to its increase of surface area and electronic structure. Yet, it is noteworthy that the increased activity of PBCO-FS, in comparison to PBCO-SG, does not surpass the relative increase of its BET surface area.

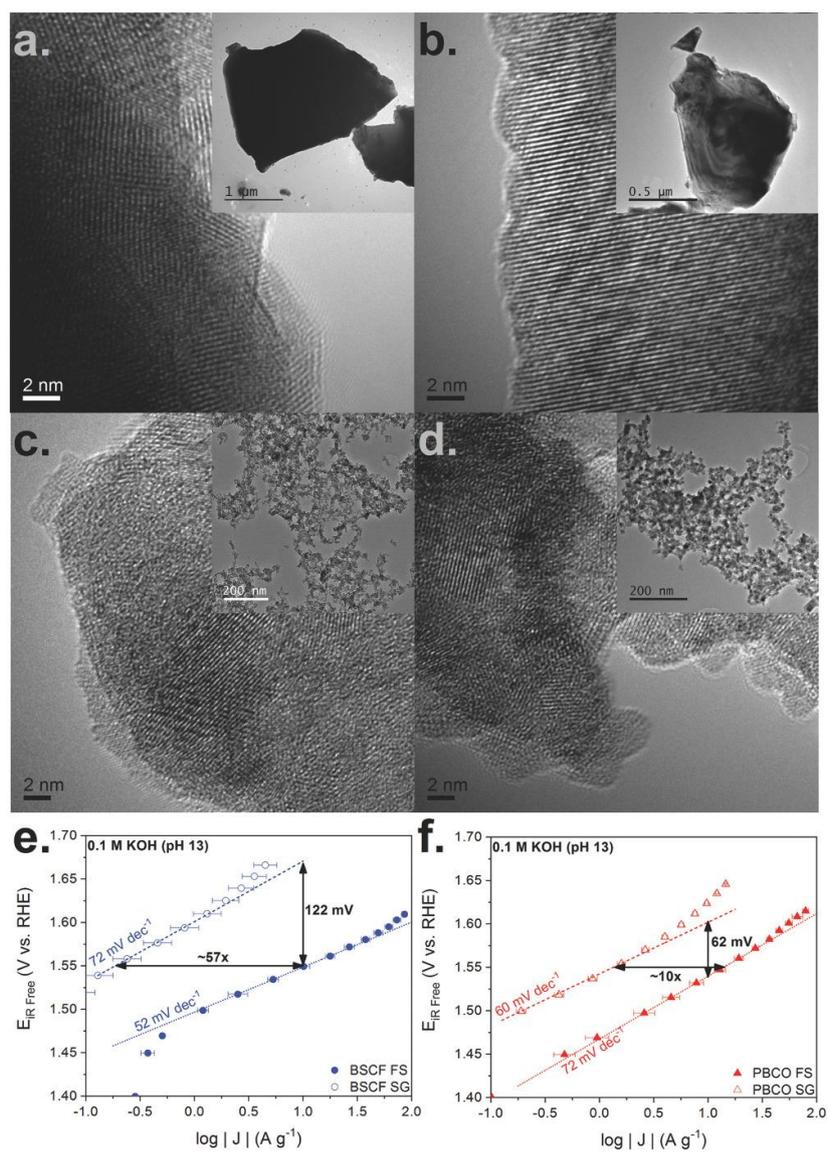

**Figure 1**
HRTEM images of a) BSCF-SG, b) PBCO-SG, c) BSCF-FS, and d) PBCO-FS. Each inset shows enlarged view of the respective catalysts. Tafel plots of e) BSCF-SG and BSCF-FS, and f) PBCO-SG and PBCO-FS in 0.1 M KOH

The direct comparison of OER activities of BSCF-FS and PBCO-FS in alkaline condition is shown in **Figure 2**a, which both perovskites show similar overpotentials at 10 A g$^{-1}$ (1.55 and 1.54 $V_{RHE}$, respectively) and current densities at 1.55 $V_{RHE}$ of 13.8 and 10.5 A g$^{-1}$, respectively, which overtake that of the precious metal catalyst (i.e., IrO$_2$ 1.59 $V_{RHE}$ at 10 A g$^{-1}$ and current density of 2.8 A g$^{-1}$ at 1.55 $V_{RHE}$) (refer to Figure S7, Supporting Information). Nevertheless, it is interesting to note the difference in their Tafel slopes suggesting that each catalyst may take different mechanistic pathways during OER with different rate determining steps.[19] All

Tafel plots are determined from chronoamperometry measurements performed after the initial 25 cycles of cyclic voltammetry (CV) in order to attain a steady-state activity of a perovskite catalyst (see Figure S3, Supporting Information). Furthermore, the difference in their underlying electrocatalytic behavior is more evident from their contrasting trends during 500 cycles of potential steps between 1.0 and 1.6 $V_{RHE}$ (Figure 2b). PBCO-FS shows a decreasing trend losing about 74% of its initial current density, while BSCF-FS reveals an increasing trend with a two-fold increase in current density at the end of the 500 cycles. In detail, the initial current densities of BSCF-FS and PBCO-FS at 1.6 $V_{RHE}$ are 5.3 and 33.4 A g$^{-1}$, respectively, which end up showing 18.8 and 8.9 A g$^{-1}$ at the end of 500th cycle, respectively. These opposing trends between BSCF-FS and PBCO-FS shown during the course of cycles reflect upon their functional stabilities, which the increasing trend of current density for BSCF-FS is appealing for practical applications.

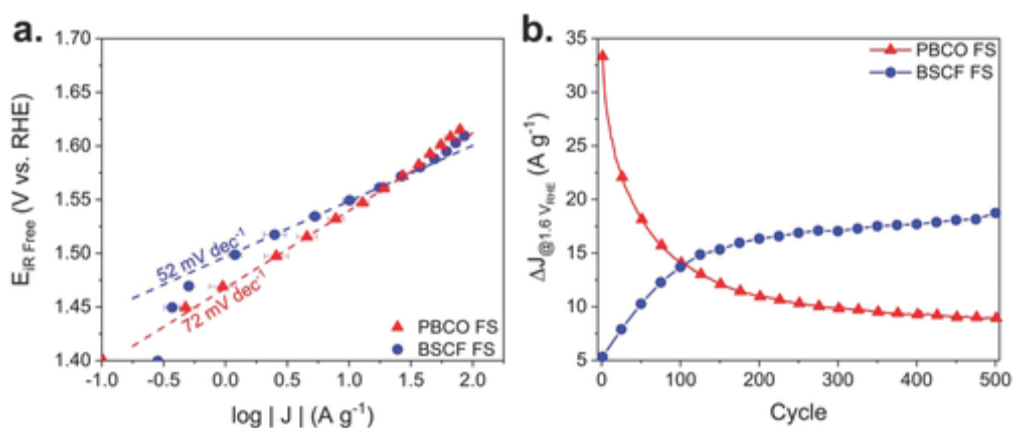

**Figure 2**
Electrochemical study comparing a) Tafel plot of OER activities and b) mass normalized current densities at every 25 cycles over 500 cycles between 1.0 and 1.6 $V_{RHE}$ of BSCF-FS and PBCO-FS in 0.1 M KOH.

As discussed in our recent study,[3] the active Co/Fe-oxy(hydroxide) species are dynamically constructed at the surface of BSCF-FS as a result of LOER likely owing to its unique capability to accommodate a high oxygen vacancy. Nevertheless, the dissolution of the A-site cations of BSCF-FS was observed during OER due to their high solubility and LOER. Meanwhile, the current density of BSCF-FS shows an increasing trend (see Figure 2b) indicating the self-construction of the active Co/Fe-O(OH) species. On the contrary, the observed loss of current density during the cycles for PBCO-FS clearly suggests the loss of its active species. As a basis of elucidating these findings, it is essential that thermodynamics of the catalyst materials under study are well comprehended. As a resolution, Pourbaix diagrams are used as an aid to understand the thermodynamic natures of these catalyst materials in aqueous electrochemical environments. Pourbaix diagrams for BSCF and PBCO (**Figure 3**a,b, respectively) are constructed with DFT calculations using 2 × 2 × 2 cubic primitive cells, which the composition step of each of A and B cation site is in multiples of

1/8 ( = 0.125, i.e., this constraint yields $Ba_{0.5}Sr_{0.5}Co_{0.75}Fe_{0.25}O_3$ rather than $Ba_{0.5}Sr_{0.5}Co_{0.8}Fe_{0.2}O_3$). Particularly, thermodynamically stable phases under alkaline conditions (pH 13) within the working potential range that corresponds to 1.0–1.7 $V_{RHE}$ (0.23–0.93 $V_{SHE}$) are to be highlighted to draw remarks on their respective functional stabilities. Referring to Figure 3a, BSCF under alkaline conditions above pH 12.6 is predicted to be meta-stable with a stability threshold of 0.5 eV atom$^{-1}$. Indeed, the A-site cations would dissociate into their aqueous phases (i.e., $Ba^{2+}_{(aq)}$ and $Sr^{2+}_{(aq)}$ in region 11) while the overall BSCF perovskite structure is retained along with Fe in its individual oxide phase within the working potential region at pH 13. Previous studies[3, 7, 20] have associated this A-site cation dissolution to the OER activity; suggesting that an amorphous layer (i.e., self-constructed metal oxy(hydroxide) layer) is created at its near-surface (consequential to the A-site cation leaching), which would increase the effective electrochemically active area and therefore render an enhanced activity. Likely, as mentioned above, this is also evident in the stability test during which BSCF-FS reveals a significant increase of mass activity in the course of the potential cycle. Furthermore, the calculated Pourbaix diagram supports our previous findings of the enhanced performance of BSCF-FS as OER catalyst based on its thermodynamic nature to retain its perovskite structure.[3]

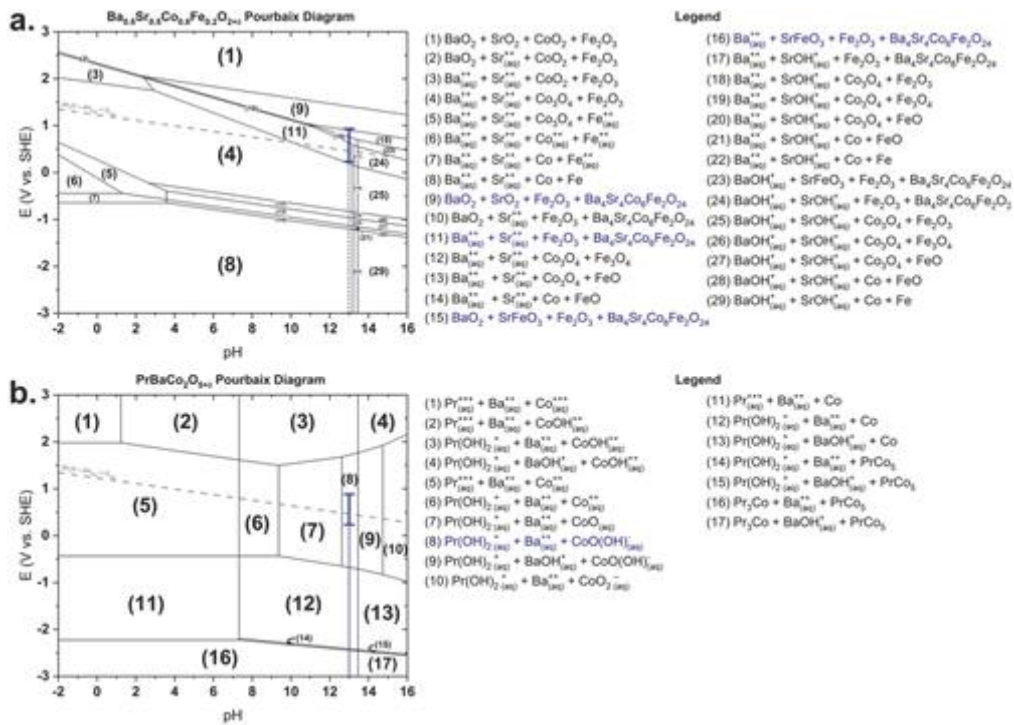

**Figure 3**

DFT-simulated Pourbaix diagrams of a) BSCF and b) PBCO. Blue bold line indicates the potential range in which chronoamperometry is carried out at pH 13.

Unlike in the case of BSCF-FS, the Pourbaix diagram of PBCO anticipates the cation disintegration of its perovskite structure, where all of the constituent cations are dissolved out into their solvated phases. Consequently, this is clearly reflected during the stability test (Figure 2b) which PBCO-FS reveals a considerable decrease in current density by 74% after cycling. Yet, a certain level of current density is still observed which may indicate that the cation dissolution is not an immediate process. A recent study[21] elaborates that although the A-site cations of PBCO—Pr and Ba—are leached out (like in the case of BSCF), PBCO tends to sustain its electrocatalytic properties longer. This was explained by relating to the defect chemistry of PBCO with the oxygen vacancy stoichiometry changing as Ba cations are leached out from the structure while Pr cations tend to remain at the near-surface. In brief, the authors explain how this degradation mechanism would not influence its short-term electrocatalytic properties. Nonetheless, the calculated Pourbaix diagram analysis suggests that PBCO would also experience B-site dissolution (region 8 of Figure 3b). In fact, our extended stability test (Figure 2b) demonstrates that electrocatalytic properties of PBCO are not perpetual.[21] Moreover, thermodynamic preference of cation dissolution of perovskite oxides during OER has been explained previously.[12] Together with the Pourbaix diagram, it can be inferred that the process of reaching an end of service life of a perovskite oxide is a process that is principally induced from cation dissolution governed by thermodynamics.

Here, one has to recognize that the dissolution mechanisms during OER are also certainly kinetic processes which obviously cannot be deduced solely from Pourbaix diagrams since they only describe phases of catalysts under equilibrium conditions. In this essence, the electrocatalysts would undergo different mechanistic pathways during the OER process depending on the reaction kinetics, which can diverge from the observations drawn from Pourbaix diagrams. Therefore, *operando* XAS was employed in order to monitor electronic and structural changes associated to the perovskite catalyst during the OER process (**Figure 4**). Figure 4 shows comparisons between PBCO-FS and BSCF-FS in their *operando* XAS measurements. Figure 4a describes positive shifts for both BSCF-FS and PBCO-FS in their Co K-edge energy levels from X-ray absorption near edge spectra (XANES) during the *operando* flow cell measurement. Considering that Co species of BSCF-FS is in a lower oxidation state than those in PBCO-FS (refer to Figure S5, Supporting Information), a greater extent of energy shift of the Co K-edge is observed for BSCF-FS (≈0.7 eV) than PBCO-FS (≈0.3 eV). In our previous study,[3] the extent of Co K-edge energy level shift is shown to be associated with the development of the OER active Co/Fe-oxy(hydroxide) layer, which is in a higher Co oxidation state than those inherent in the perovskite oxide structure. This is also evident in Figure 4b as BSCF-FS shows a rapid shift of Co K-edge energy when polarized into the oxygen evolution regime (above 1.4 $V_{RHE}$). In the case of PBCO-FS, referring to Figure 4b, the Co K-edge energy shift is revealed to be consistently increasing even at potentials below the oxygen evolution regime, which is indicative of oxidation of Co species based on the anticipated dissolution described by its Pourbaix diagram (Figure 3b, region 8). In relevance to the stability test (Figure 2b), this complies with the observed decreasing trend of activity upon potential cycling, which can be attributed to the cation disintegration from the perovskite structure of PBCO. In summary, the DFT calculated Pourbaix diagram anticipates meta-stability of BSCF, which then would enable BSCF to retain its perovskite structure providing substrate surface for the self-construction of active oxy(hydroxide) layer and LOER. In contrast, the increase of Co oxidation state in PBCO at potentials below the onset of OER suggests chemical dissolution of its cations regardless to the event of LOER as governed by thermodynamics.

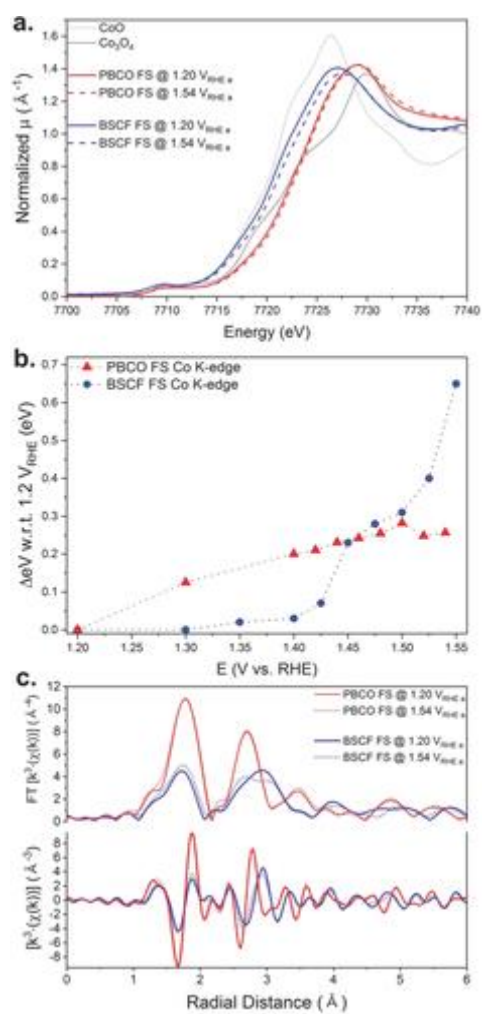

**Figure 4**

*Operando* XAS study results of PBCO-FS and BSCF-FS; a) comparison of XANES of Co K-edge at 1.2 and 1.54 $V_{RHE}$, b) edge shift measured with respect to the edge position at 1.2 $V_{RHE}$ at each potential during operando flow cell test, and c) comparison of FT-FT-EXAFS spectra at 1.2 and 1.54 $V_{RHE}$.

Figure 4c shows the comparison of Fourier-transformed (FT) k³-weighted Co K-edge extended X-ray absorption fine structure (EXAFS) spectra of PBCO-FS and BSCF-FS. The EXAFS spectra of BSCF-FS spectra reveals changes of peaks associated with Co–O and Co–Co coordination shells each located at radial distances of ≈ 1.9 and ≈2.6–3.0 Å, respectively. First, the amplitude of the Co–O peak is increased indicating that the absorbing atom is bounded by more oxygen atoms at that radial distance. This is further supported by the observed positive shift in the Co K-edge from its *operando* XANES (Figure 4b). Meanwhile, the second peak associated with the Co–Co coordination of BSCF-FS undergoes significant changes during the anodic polarization. At 1.54 $V_{RHE}$, the Co–Co peak clearly shows two peaks at slightly different radial distances of which the earlier peak (≈2.8 Å) has evolved to become more distinct. Initially at 1.2 $V_{RHE}$, the observed FT-EXAFS spectrum of BSCF-FS reveals mainly a single peak for its Co–Co coordination shell at ≈3.0 Å, which is reduced in its amplitude as the applied potential is increased into the oxygen evolution regime. This single Co–Co peak is attributed to edge-shared polyhedron which is from having a highly oxygen deficient perovskite oxide.[22] The earlier Co–Co peak at ≈2.8 Å becomes apparent as polarization continues which is ascribable to Co–Co coordination shell of Co/Fe–O(OH).[23] This is known to be the key feature behind its high OER activity.[3, 24] As the Co/Fe-oxy(hydroxide) is in a higher oxidation state (≈3) than that of the Co species of BSCF-FS perovskite structure (≈2.2), the increase of the Co oxidation state manifested by the shift of its Co K-edge in XANES (Figure 4b) describes the construction of the active oxy(hydroxide) layers.

Regarding the *operando* FT-EXAFS of PBCO-FS, however, the cation dissolution is not evidently highlighted when comparing the spectra observed at the initial and the final potentials of the anodic polarization as they do not show significant changes (see Figure 4c). These insignificant changes can be elucidated by referring back to its Pourbaix diagram. Based on the region 8 of Figure 3b, it is anticipated that Ba and Pr of the perovskite structure would be dissolved into their respective aqueous phases (i.e., $Ba^{++}_{(aq)}$ and $Pr(OH)_2^+{}_{(aq)}$, respectively), while Co would be oxidized to $CoO(OH)^-_{(aq)}$.[21] Unfortunately, $CoO(OH)^-_{(aq)}$, the oxidized product of Co from the perovskite structure, exhibits a similar Co–Co radial distance as that of edge-sharing Co octahedron within perovskite structure itself, which are generally induced by a high oxygen deficiency within the perovskite oxide. By the structural analysis obtained by XRD and neutron diffraction on BSCF and PBCO samples (Table S2, Supporting Information) the Co–Co distances of the edge-shared polyhedron are estimated to be 2.793 and 2.741 Å for BSCF-FS and PBCO-FS, respectively. Notice that the Co–Co distances of edge-sharing polyhedron of PBCO-FS is shorter than that of BSCF by ≈0.05 Å. Coincidentally, the estimated Co–Co distance of edge-sharing polyhedron of PBCO are close to that of CoO(OH) (≈2.5-2.7 Å).[23] Therefore, the peak of the Co–Co coordination shell in FT-EXAFS of PBCO-FS could be attributed both to Co–Co edge-shared polyhedron scattering and to Co–Co scattering from Co–O(OH) species (see Figure S6, Supporting Information). Consequently, only limited information can be extracted from its FT-EXAFS spectrum regarding the construction of the active CoO(OH) and/or the dissolution of Co from the perovskite. Based on this, even though the dissolution of PBCO is eventuated, it can only be speculated that the dissolution is not an immediate process owing to the hindrance of Co cation dissolution of PBCO during OER/LOER by the Pr hydroxyl phase formed at the surface as suggested in Bick et al.[21] Furthermore, in the case of BSCF, Fe-doped Co–O(OH) is formed owing to the Fe content in its perovskite structure, while only CoO(OH) can be possibly formed at the surface of PBCO. It is prevalent that Fe conveys beneficial effects to the Co-based catalyst with respect to their OER activity.[24, 25] Therefore, the enhanced OER activity of BSCF would attribute to the presence of Fe with Co, while in the case of PBCO, in the absence of Fe, would continue in its degradation process as a less active oxy(hydroxide) layer is formed.

Additional to the fundamental understanding of thermodynamics and the operando changes of the catalyst during the OER process, the practical performances of BSCF-FS and PBCO-FS are corroborated by adopting these perovskites as anodic electrode materials in a full AEMWE. **Figure 5** shows AEMWE performances for membrane electrode assemblies (MEAs) having BSCF-FS, PBCO-FS, and commercial $IrO_2$ as anodic electrode under actual operating conditions. Figure 5a,b presents polarization curves and voltage versus time behaviors at steady-state current of 500 mA $cm^{-2}$, respectively. Conforming to the trend observed in the rotating disk electrode (RDE) study, both perovskites revealed higher activities than commercial $IrO_2$, among which the cell with BSCF-FS-based MEA demonstrates the highest activity (Figure 5a). It is noteworthy that the performance of BSCF-FS in AEMWE improves over time when the galvanodynamic test is repeated (refer to Figure S8, Supporting Information). Additionally, Figure 5b shows that the cell with BSCF-FS based MEA demonstrates the lowest and the most stable cell voltage over time at the steady-state current of 500 mA $cm^{-2}$. On the other hand, although the cell with PBCO-FS-based MEA shows a similar activity as the commercial $IrO_2$ in the galvanodynmaic test (Figure 5a), a higher cell voltage was required than the cell with BSCF-FS-based MEA for the steady-state current of 500 mA $cm^{-2}$.

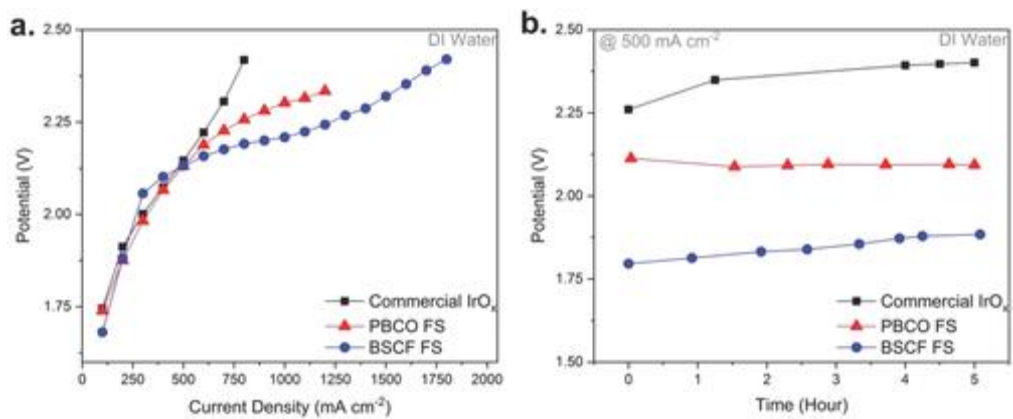

**Figure 5**

Performance comparison of BSCF-FS, PBCO-FS, and IrO$_2$ anode OER catalyst in a technical AEMWE. a) Polarization curves and b) voltage versus time at steady-state current density of 500 mA cm$^{-2}$ obtained for MEAs having BSCF-FS, PBCO-FS, and IrO$_2$ as anodic electrode.

# 3 Conclusion

In this work, we assess the state-of-the-art perovskite catalysts—PBCO and BSCF—from a thermodynamic perspective that evaluates their OER performances for AEMWE. In this context, we first report remarkably enhanced OER performances of PBCO-FS and BSCF-FS in alkaline electrolyte, which attribute to their uniformly distributed nanoparticles and a capability to facilitate changes in their Co oxidation states imposed by the FS synthesis method. Although similar current densities were attained for both perovskite catalysts (13 and 11 A g$^{-1}$ for PBCO-FS and BSCF-FS, respectively), their current density show diverging trends during the stability test, with BSCF-FS showing an exceptional long-term performance.

In order to elucidate the observed stabilities and activities of PBCO and BSCF, DFT-calculated Pourbaix diagrams are constructed. First, the Pourbaix diagram describes that the BSCF perovskite structure is thermodynamically meta-stable in alkaline electrolyte above pH 12.6 within the OER potential range; its A-site cations (Ba and Sr) are partially dissolved while the initial perovskite structure is retained. On this basis, the high-mass activity of BSCF-FS in the alkaline electrolyte, therefore, can be thermodynamically rationalized. This explanation is consistently valid for the previously reported behavior of BSCF.[**3**] In the case of PBCO, a decreasing trend of current density was observed which confirmed its thermodynamic instability as anticipated by the Pourbaix diagram.

From the above findings, important remarks are drawn relating thermodynamically stable phases of perovskite catalysts to their respective functional stability. First, the loss of initial activity during the stability test corresponds to the loss of perovskite structure as a result of A- and B-site cation dissolutions. Exceptionally, BSCF is considered to be meta-stable while allowing its perovskite structure to serve as the substrate for the active oxy(hydroxide) to self-construct. This has been corroboratively confirmed by the results gathered from *operando* XAS study, which the comparison of FT-EXAFS spectra revealed evident growth of the peak associated with the active CoO(OH) species. This is further supported by the increase of oxidation state recorded from reading the energy shift of Co K-edge during operando study.

On the other hand, in the case of PBCO-FS, the overlapping of Co–Co radial distance between those of CoO(OH) species and edge-sharing Co polyhedron veils the reading of geometric structural changes of the perovskite catalyst. However, an increase of the Co oxidation state of PBCO-FS was manifested below the oxygen evolution potential regime, suggesting the occurrence of chemical dissolution processes. Overall, insights into the dissolution mechanism of PBCO-FS is vital in order to completely understand its functional stability and activity towards OER as these processes are intertwined.

Even though the cation dissolution is driven by thermodynamic forces, one should keep in mind that the dissolution is not an immediate process but occurs at a kinetically controlled rate. Therefore, kinetic parameters need to be clearly identified in order to come to a full understanding of OER activity of a perovskite oxide catalyst. Overall, the thermodynamically stable phases illustrated by a Pourbaix diagram show the most stable phase that a perovskite would transform into, but does not describe the transformation pathway. It is along this pathway that the OER activity of a perovskite catalyst is transpired, from which the understanding of its electrocatalytic behavior should hence be referred.

Based on the knowledge obtained from this investigation, the degradation mechanism of perovskite oxides is the key to fully understand its functional stability during the electrochemical reactions, which then can be also related to its activity. The presented approach of unraveling the relationships between functional stability, structure, and activity of perovskite oxides during OER will pave the way for optimizing an active perovskite catalyst for technical applications.

# 4 Experimental Section

*Material Syntheses*: For the preparation of the PBCO precursor solution, stoichiometric amounts of praseodymium oxide ($Pr_6O_{11}$, 99.9%, Auer Remy) were dissolved in water and 5 vol.% nitric acid (70%, Alfa Aesar) at 80 °C, before mixing with stoichiometric amounts of barium carbonate (≥99.0%, Sigma–Aldrich) and cobalt nitrate hexahydrate (99.9%, Auer Remy), acetic acid (≥99.0%, Fluka) in 25 vol.%, and N,N-Dimethylformamide (≥99.8%, Roth) in 45 vol.%. The total metal concentration in solution was fixed to 0.1 M.

The BSCF precursor solution was prepared by dissolving stoichiometric amounts of barium carbonate (≥99.0%, Sigma–Aldrich), strontium nitrate (≥98%, Sigma–Aldrich), cobalt nitrate hexahydrate (99.9%, Auer Remy), and iron nitrate non-ahydrate (≥98%, Sigma–Aldrich) in a solution of de-ionized water (DIW) and acetic acid (≥99.0%, Fluka) 25 vol. %. The total metal concentration in solution was fixed to 0.2 M.

For all the syntheses, the flow rate of dispersing oxygen was fixed at 35 mL min$^{-1}$, while the flow rates of the combustion gases were fixed at 17 and 13 mL min$^{-1}$ for the mixture of oxygen and acetylene.

*Material Characterizations*: Phase characterization of prepared materials was performed using powder XRD (Bruker D8 system in Bragg–Brentano geometry, Cu K-α radiation (λ = 0.15418 nm)). Specific surface area was calculated by BET analysis of $N_2$ adsorption/desorption isotherms (AUROSORB-1, Quantachrome). TEM and energy dispersive X-ray spectroscopy (TECNAI F30 operated at 300 kV) were used to study the surface morphology and composition of the prepared materials.

*Electrochemical Characterization*: The electrochemical activities of the prepared catalysts were evaluated in a standard three-electrode electrochemical cell using the thin-film RDE methodology. The setup for OER and CV consists of a potentiostat (Biologic VMP-300) and a rotation speed controlled motor (Pine Instrument Co., AFMSRCE). All the electrochemical measurements were performed at the standard room temperature using a reversible hydrogen electrode (RHE) as the reference electrode in 0.1 M KOH. A piece of gold mesh was used as the counter electrode separated via a salt bridge. A homemade Teflon cell was used to contain the electrolyte with the working electrode immersed under potential control. A porous thin film electrode was prepared by drop-casting a catalyst ink on a polished glassy carbon electrode (5 mm OD/0.196 cm$^2$).**26** The catalyst ink was prepared from a catalyst suspension made from sonicating (Bandelin, RM 16 UH, 300 Weff, 40 kHz) 10 mg of catalyst in a solution mixture of 4 mL isopropyl alcohol and 1 mL of Milli-Q water (ELGA, PURELAB Chorus), and 20 µL of Nafion (Sigma–Aldrich, 5 wt. %). The same method of preparation of the working electrode was used for all the samples. The 0.1 M KOH electrolyte was prepared by dissolving KOH pellets (Sigma–Aldrich, 99.99%) in Milli-Q water. Initially, 25 reverse potentiostatic sweeps of CV were performed in synthetic air-saturated electrolyte from 1.0 to 1.7 $V_{RHE}$ at a scan rate of 10 mV s$^{-1}$. Subsequently, chronoamperometric measurements were carried out holding each potential step for 30 s to obtain steady-state currents in the potential range from 1.2 to 1.7 $V_{RHE}$ while rotating the working electrode at 900 rpm. Tafel plots were constructed from the resulting polarization curves for all materials under study for an effective comparison of OER activities. The potential stability of catalyst materials was conducted using the same setup; alternating electrode potential between 1.0 and 1.6 $V_{RHE}$ for 500 times holding for 10 s each to reach steady state at each potential. Five cycles of CV and electrochemical impedance spectroscopy (EIS) were carried out after every 100 cycles. All measured currents were normalized by the mass of catalyst loading, and potentials were corrected for the ohmic drop obtained from EIS.

*Alkaline Exchange Membrane Water Electrolyzer Test*: For the MEA, each anodic catalyst (BSCF-FS, PBCO-FS, and commercial IrO$_2$) and cathode catalyst (Pt) were prepared by mixing the catalyst with water, isopropanol, and ionomer into an ink.**27** The ink also consisted of a non-functional binder in addition to the ionomer supplied by Tokuyama. The ink was made in several stages, with stirring in between each stage. The catalyst ink mixture was then sprayed onto the gas diffusion layer (GDL, Tokuyama) substrate using an airbrush to form gas diffusion electrodes (GDE). GDE's generally contained total catalyst loading of 3 mg cm$^{-2}$, and GDE composition was measured by mass change of the GDL. The cathode GDL's were carbon paper, while the anode GDL's were made from a titanium porous transport layer, which titanium oxide was purchased from Umicore AG & Co KG. Electrolysis testing was conducted using a modified 25 cm$^2$ Fuel Cell Technologies cell, where the graphite flow field on the oxygen evolution side was replaced with a stainless steel parallel channel flow field for stability at high electrolysis potentials. The cell was assembled by placing the appropriate GDE on each side of the anion exchange membrane (Tokuyama A201). The MEA was assembled using Tefzel gaskets and by applying 4.52 Nm torque to the bolts. GDE's and membrane were exchanged in 0.5 M sodium hydroxide solution for 1 h prior to testing. The cell was supplied either with DIW on the anode side of the cell only. The cathode side was not pressurized and no sweep gas was used. The temperature was maintained at 50 °C during operation. After the cell reached the operating temperature, galvanodynamic test was carried out to obtain a polarization curve, and then a galvanostatic test was carried out at 500 mA cm$^{-2}$ to assess initial stability.

*Operando Flow Cell Test*: For XAS measurements, catalyst powders were dispersed in a mixture of isopropanol and Milli-Q water in the equal ratio sonicated for 30 min, and spray coated on Kapton film. XAS spectra at the Co and Fe K-edge were recorded at the SuperXAS beamline of the Swiss Light Source (PSI, Villigen, Switzerland). The incident photon beam provided by a 2.9 T superbend magnet source was collimated by a Si-coated mirror at 2.85 mRad (which also served to reject higher harmonics) and subsequently monochromatized by an Si (111) channel-cut monochromator. Rh-coated toroidal mirror at 2.5 mRad to focus the X-ray beam to a spot size of Ø 1 mm on the sample position. The SuperXAS beamline[28] allowed for the rapid collection of 120 spectra during a measurement time of 60 s (QEXAFS mode), which were then averaged. The spectra of samples were collected in transmission mode using $N_2$ filled ionization chambers, where a Co foil between the second and third ionization chambers served to calibrate and align all spectra. EXAFS spectra were analyzed using the Demeter software package,[29] which included background subtraction, energy calibration (based on the simultaneously measured Co or Fe reference foil), and edge step normalization. The resulting spectra were converted to the photoelectron wavevector $k$ (in units Å$^{-1}$) by assigning the photoelectron energy origin, $E_0$, corresponding to $k = 0$, to the first inflection point. The resulting $\chi(k)$ functions were weighted with $k^3$ to compensate for the dampening of the EXAFS amplitude with increasing $k$. These $\chi(k)$ functions were FT over 3–12 Å$^{-1}$.

The electrode materials were spray coated at the center of a Kapton film. Black Pearls (Cabot Corp.) were used as the counter electrode material. A silver chloride electrode ($Ag^+$/AgCl) (Hugo Sachs Elektronik) was used as the reference electrode. During electrochemical testing, the electrolyte was drawn into the cell and collected in a syringe at the flow rate of 0.1 mL min$^{-1}$. The chronoamperometry measurement was carried out holding for 2 min at each increasing potential step from 1.2 to 1.54 $V_{RHE}$, and again during the reverse sequence back to 1.2 $V_{RHE}$. At each potential step, transmission XAS spectra at Co K-edge were collected simultaneously for 1 min. The position of the Co K-edge is determined at the half edge-step intensity. The *operando* XAS flow cell was described in detail in our previous study.[11]

*Density Functional Theory—Pourbaix Diagrams*: The phase stability and stability in aqueous system of each perovskite was investigated using quantum mechanical simulations in the framework of DFT. All the structures considered have been fully relaxed using the Quantum ESPRESSO package,[30] choosing PBEsol as the exchange-correlation functional[31] and pseudopotentials from the Standard Solid State Pseudopotential library (SSSP accuracy).[32] Hubbard U+ correction of 4 eV is used and applied to Co, Fe, and Pr elements. The energies obtained from DFT have then been used to draw the Pourbaix diagrams by means of the phase diagram module implemented in the Atomic Simulation Environment.[33]

With these DFT Pourbaix diagrams we investigated the stability in water as a function of pH and electrochemical potential. We compare calculated DFT energies of all the possible solid phases in which the perovskites can separate (as enumerated in the Materials Project

database**34**) with experimentally measured dissolution energies for the dissolved phases.**35** Details of this method are available elsewhere in the literatures.**36**

## Acknowledgements

The authors gratefully acknowledge the Swiss National Science Foundation through its Ambizione Program and the NCCR Marvel, CCEM through the project RENERG2, the Swiss Competence Center for Energy Research (SCCER) Heat & Electricity Storage through Innosuisse, Switzerland, and Paul Scherrer Institute for financial contributions to this work, respectively.

## Conflict of Interest

The authors declare no conflict of interest.

# Supporting Information for Adv. Funct. Mater., DOI: 10.1002/adfm.201804355

Highly Active Nanoperovskite Catalysts for Oxygen Evolution Reaction: Insights into Activity and Stability of Ba0.5Sr0.5Co0.8Fe0.2O2+δ and PrBaCo2O5+δ

Bae-Jung Kim,* Xi Cheng, Daniel F. Abbott, Emiliana Fabbri,* Francesco Bozza, Thomas Graule, Ivano E. Castelli, Luke Wiles, Nemanja Danilovic, Katherine E. Ayers, Nicola Marzari, and Thomas J. Schmid

## S1. X-Ray Diffraction Analysis

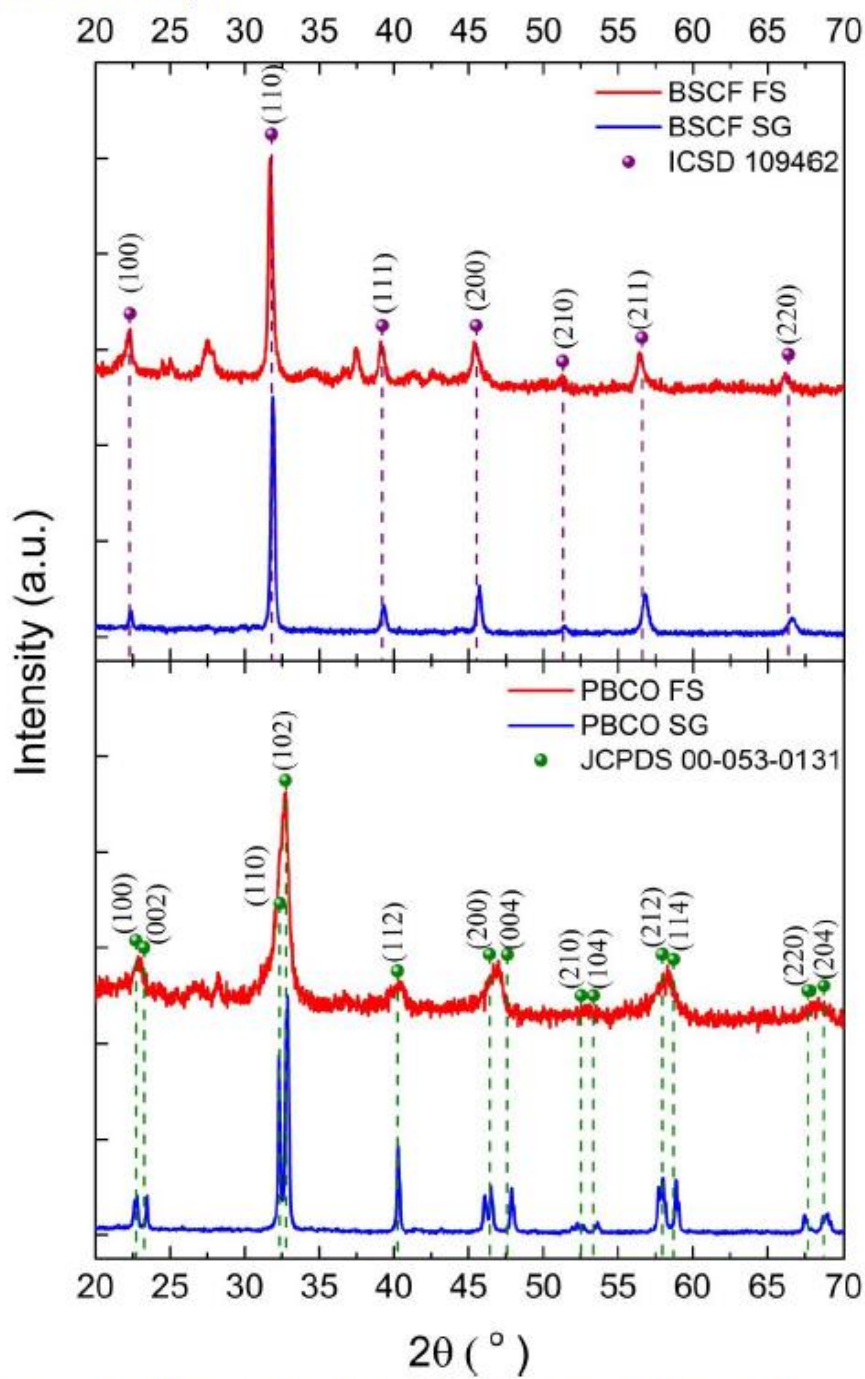

Figure S1. Comparisons of XRD patterns of a) BSCF and b) PBCO from sol-gel (SG) and flame spray method (FS).

## S2. Brunauer, Emmett, and Teller (BET) Surface Area

Table S1. BET surface areas of all catalyst materials studied

| Catalyst Material | | BET Surface Area ($m^2 \cdot g^{-1}$) |
|---|---|---|
| BSCF | SG | 2.7 |
| | FS | 28.3 |
| PBCO | SG | 2.6 |
| | FS | 43.8 |
| $Co_3O_4$ | Alfa Aesar | 1.4 |

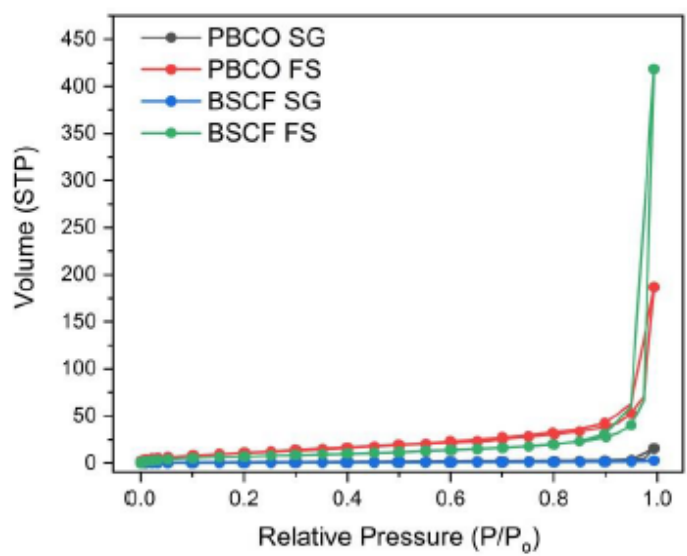

Figure S2. BET Isotherm curves of the catalysts.

## S3. Electrochemical Analysis
Cyclic Voltammetry

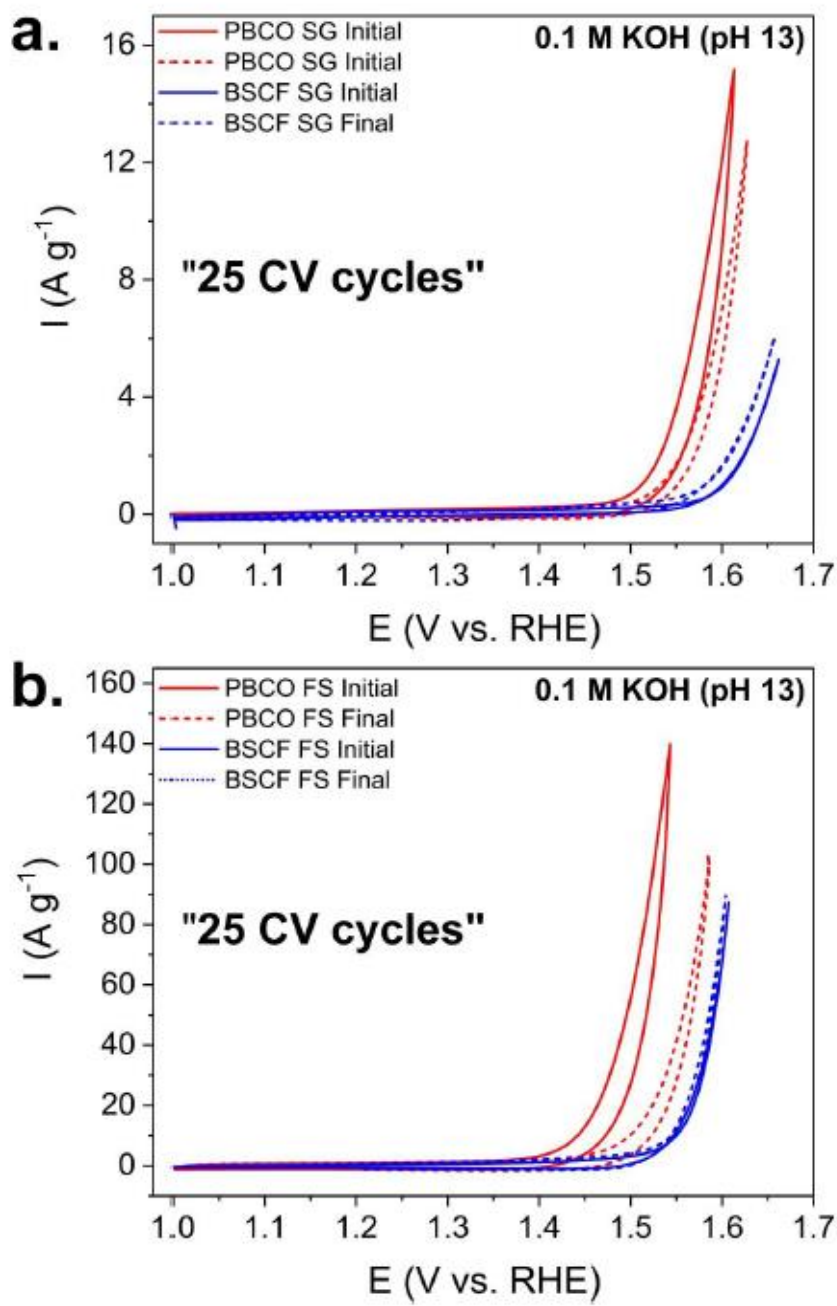

Figure S3. Comparisons of 25 CV cycles prior to chronoamperometry measurement in 0.1 M KOH (pH 13) between BSCF and PBCO prepared by a) sol-gel synthesis and b) flame spray method.

## S4. XAS Study of Perovskite Catalysts

### S4.1 Prepared by Sol-gel and Flame Spray Synthesis

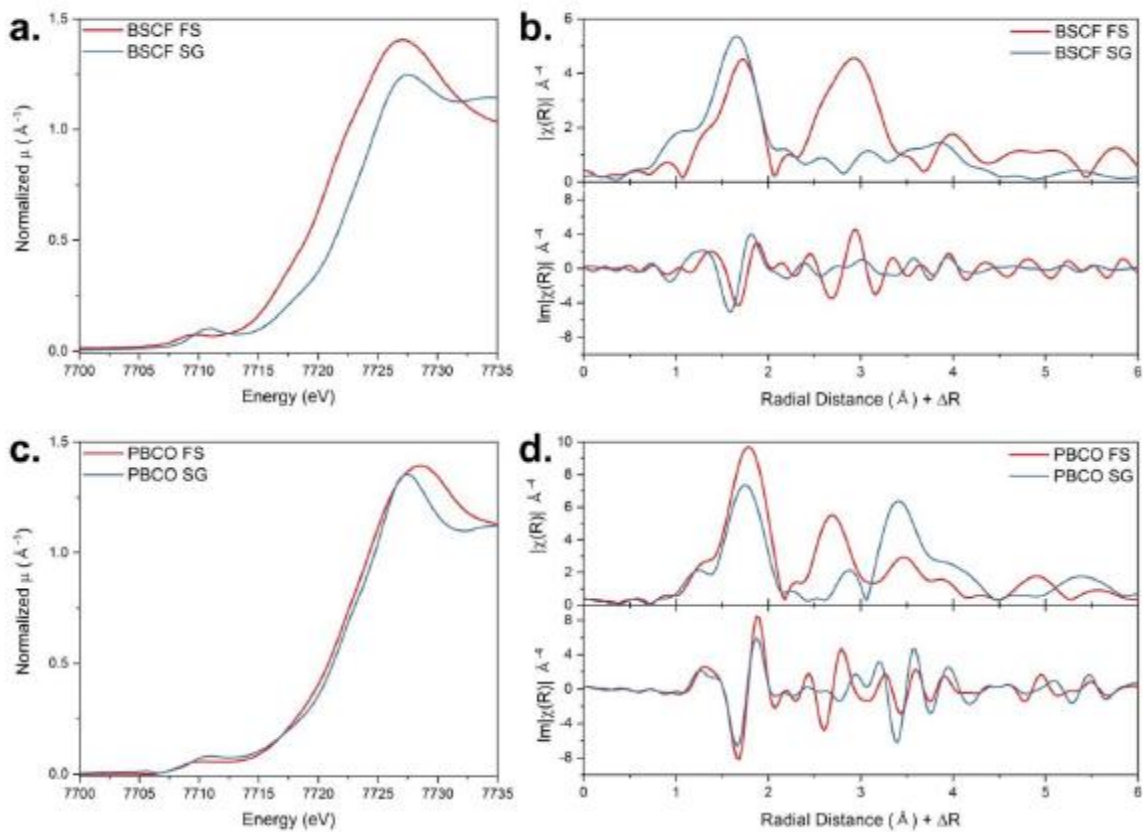

**Figure S4.** Comparison of X-Ray absorption near edge spectra (XANES) and extended X-ray absorption fine structure (EXAFS) spectra of the as-prepared catalysts from sol-gel (SG) and flame spray (FS) syntheses: a – b) XANES and EXAFS, respectively, of BSCF-SG and BSCF-FS; c – d) XANES and EXAFS, respectively, of PBCO-SG and PBCO-FS.

The higher Tafel slope can be explained by various factors related to the physico-chemical differences of the catalysts produce by sol-gel and flame spray synthesis. First, **Figure S4a** and c clearly show that the Co oxidation states show differences for BSCF-FS and PBCO-FS, respectively, from the two synthesis methods. In regards to the peak energy positions, the flame spray synthesis for both BSCF and PBCO shows lower Co oxidation states than those prepared by sol-gel method. In case of PBCO, the difference in the Co oxidation state between FS and SG is not as distinct as in the case of BSCF since the balance of oxidation state of the perovskite structure must be maintained. Therefore, Co in PBCO would have less flexibility in its oxidation state to exist in a lower oxidation state; while for BSCF, the presence of Fe would allow Co to be in a lower oxidation state.

Furthermore, referring to Figure S4b and d, EXAFS of both BSCF and PBCO show different patterns after the first peak ascribable to the Co–O coordination shell. These differences in EXAFS pattern signify that each perovskite catalyst prepared by FS has different radial distances with its surrounding atom.

For example, in the case of BSCF-SG, the second peak is attributable to its Co–Co coordination shell which appears at around 3.4 Å. Meanwhile, the peak position of Co–Co coordination shell of BSCF FS shows that it is at a closer radial distance (~ 2.8 – 2.9 Å) indicating a closer bond distances with other neighboring Co atoms. This could be partially from having the Co in a lower oxidation state, and which would also lead to affect the arrangement in sharing of Co atoms as to share edges of the Co octahedron rather than to be shared via its Co corner. The differences in Co octahedron network are also made evident in the pre-edge features, which is known to be associated with the distortion of transition metal ion octahedron and distribution of the neighboring cations.[1] Therefore, the observed differences in their pre-edge position of SG and FS products would attribute to having a different quantity of oxygen vacancy; henceforth, they would exhibit different electronic configurations especially affecting their 3d shell.

In case of PBCO, Figure S4c clearly shows that PBCO-FS has a slightly lower Co oxidation state than that of PBCO-SG. In addition, as similar to the case of BSCF, EXAFS of PBCO-FS and PBCO-SG show different peak patterns after the first Co–O peak (Figure S4d). The second peak of the EXAFS that is ascribable to Co-Co coordination shell of PBCO-FS is revealed to be at a closer radial distance (~2.7 Å) than that of PBCO-SG (~3.4 Å). As similar to the case of BSCF-FS, the shorter radial distance of Co–Co coordination shell suggests the presence of edge-sharing Co octahedron in PBCO-FS induced by a high oxygen deficiency; reorganize its polyhedral formation to a more stable network by rearranging with neighboring polyhedra to share their edges.[2]

Considering the differences observed from the XAS study, the catalysts produced by flame spray synthesis differ from those prepared by sol-gel not only in their particle sizes but also in their Co oxidation state and coordination of surrounding atoms. Therefore, these physico-chemical characteristic differences observed for these catalysts obtained by two synthesis methods would thereby affect their OER mechanistic processes during the polarization. Nevertheless, the Tafel slope interpretation for the oxygen evolution reaction is extremely complex, and indeed the studies dealing with OER Tafel slope analysis are uncommon. Without a solid literature, we could only speculate about the influence of the catalyst physico-chemical properties on the Tafel slope.

## S4.3 EXAFS of as-prepared Perovskite Oxides

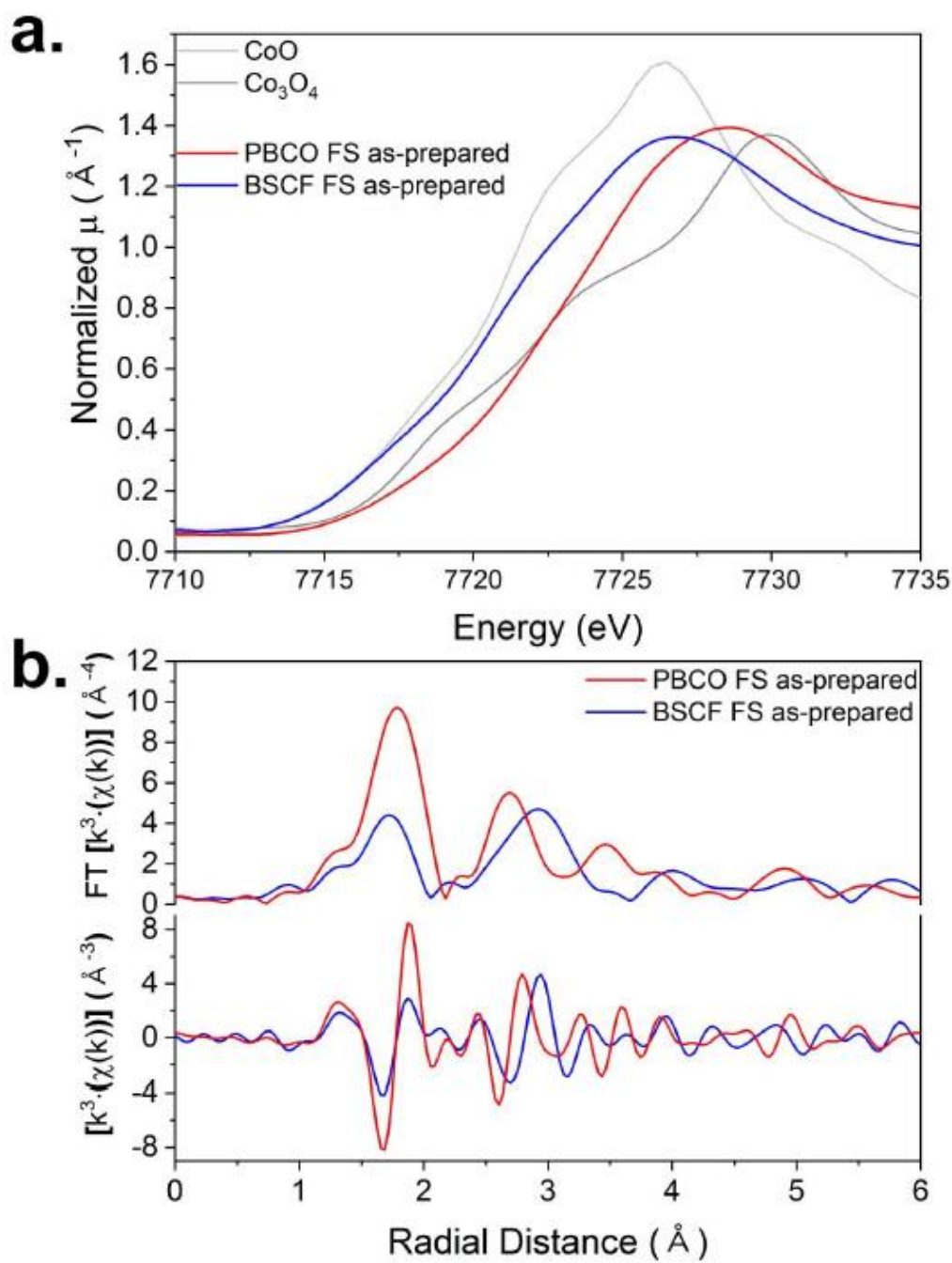

**Figure S5.** X-ray absorption spectroscopy study of the as-prepared of PBCO-FS and BSCF-FS; comparison of a) X-ray absorption near edge spectra and b) extended X-ray absorption fine structure spectra.

## S4.2 PBCO-FS Lattice Parameter and EXAFS

Table S2. Lattice parameters calculated from either X-ray diffractions or neutron diffraction analysis. From this, Co–Co of edge shared polyhedra is calculated.

| Perovskite Oxide | Methods | $a$ Co-Co/Fe corner-shared octahedra (Å) | $\frac{a}{2\sqrt{2}}$ edge-shared polyhedra (Å) |
|---|---|---|---|
| BSCF | Neutron[3] | 3.9497 | 2.793 |
| PBCO | X-ray | 3.8766 | 2.741 |

Given that the cation ordering within PBCO and BSCF is distinctively different – ordered and random, respectively – the geometric structure of each perovskite oxide would be unique. Thereby, their EXAFS of its radial distance from the absorbing Co to surrounding atoms would be notably different. Lattice parameters of each perovskites in our study have been determined using Rietveld refinement from XRD and neutron diffraction analysis for PBCO-FS and BSCF-FS, respectively (Table S2). Note that the Rietveld refinement was carried on XRD of PBCO-FS instead of neutron diffraction due to refinement limitations arising from the broadness of their neutron diffraction peaks (refer to our previous work[3] for the neutron diffraction data of BSCF-FS). Refer to the index at the end of S4.2 for the fittings.

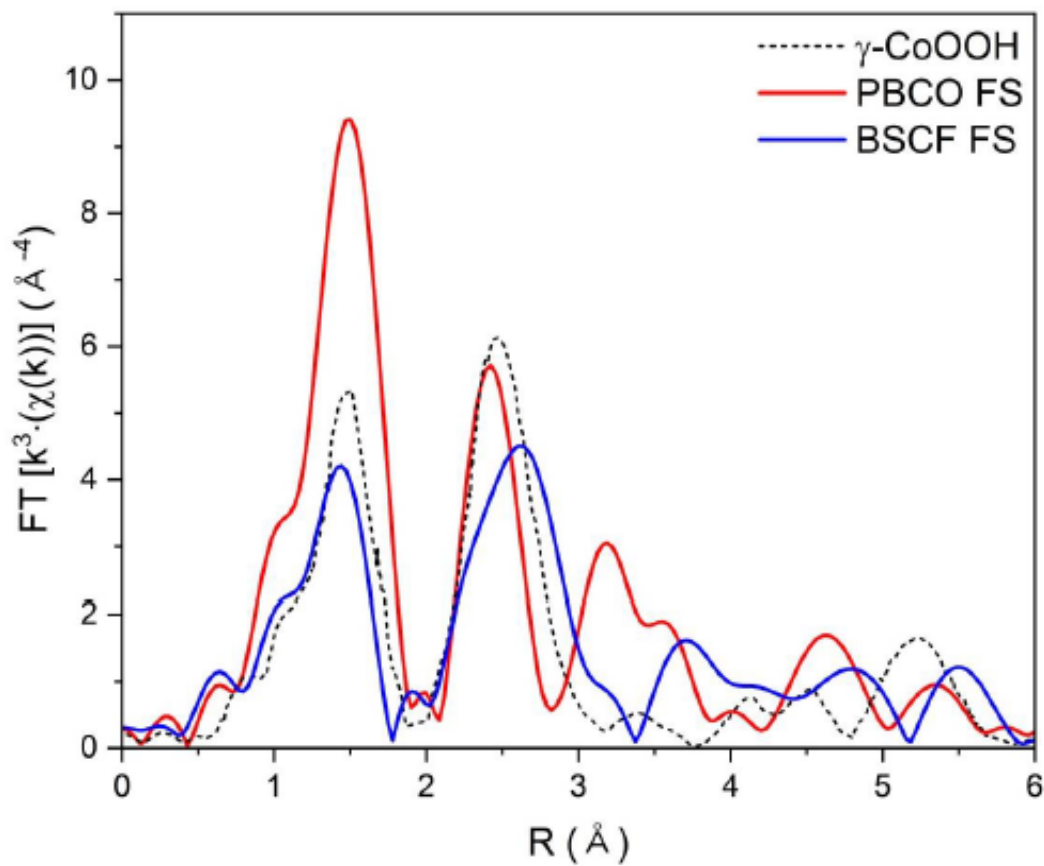

**Figure S6.** Comparison of Fourier transformed $k^3$-weighted EXAFS at Co K-edge of γ-CoOOH[4] with as-prepared PBCO-FS and BSCF-FS.

As mentioned in the above section, the second peak of the EXAFS that is associated with Co-Co coordination shell of the catalysts from flame spray synthesis is ascribable to the edge-sharing Co octahedron. The edge-sharing polyhedrons are generally induced by a high oxygen deficiency within the perovskite oxide.[2, 5] The Co–Co distances of the edge-shared polyhedrons are estimated for each perovskite in Table S2. Notice that the Co–Co distances of edge-sharing polyhedron of PBCO-FS is shorter than that of BSCF by ~0.05 Å. Coincidentally, the estimated Co–Co distance of edge-sharing polyhedron of PBCO are close to that of CoO(OH) (~2.5-2.7 Å).[4, 6] Therefore, the peak of the Co–Co coordination shell in FT-EXAFS of PBCO-FS could be attributed both to Co-Co edge-shared polyhedron scattering and to Co-Co scattering from Co-O(OH) species, which therefore limits the observable changes under *operando* conditions. Given the lattice parameter of BSCF (Table S2), on the other hand, the Co–Co bond distance of edge-sharing polyhedron is larger than that of Co-O(OH) making the two species distinguishable (refer to **Figure S6**).

*Index*

$Ba_{0.5}Sr_{0.5}Co_{0.8}Fe_{0.2}O_{2+x}$  Pm-3m 300K (space group 221)
a = 3.9497 Å    V = 61.6158 Å³

| Atom | Occupancy | x | y | z | Thermal Coefficient B (Å³) |
|---|---|---|---|---|---|
| Ba | 0.5 | 0 | 0 | 0 | 1.04 |
| Sr | 0.5 | 0 | 0 | 0 | 1.04 |
| Co | 0.8 | ½ | ½ | ½ | 1.45 |
| Fe | 0.2 | ½ | ½ | ½ | 1.45 |
| O  | 2.255 | ½ | ½ | 0 | 2.45 |

Rwp : 4.05 %
Gof : 2.05
$R_{Bragg}$ : 11.4 %

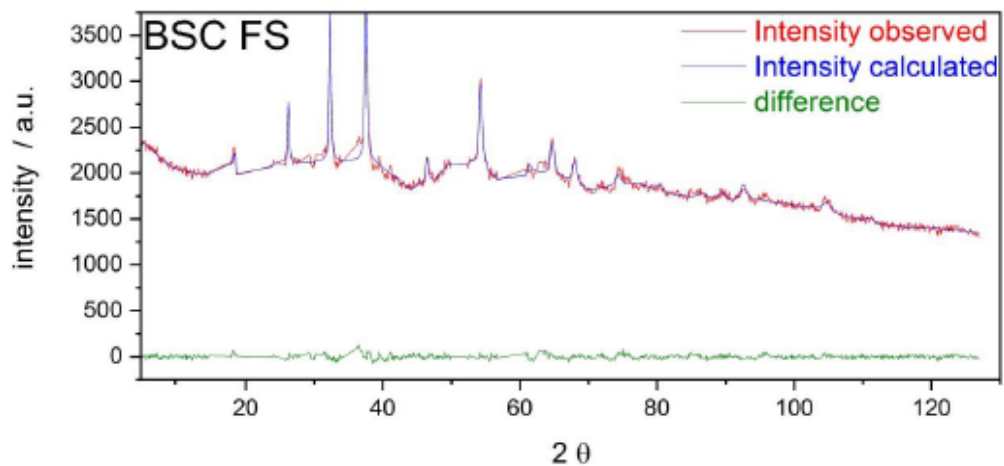

**PrBaCo$_2$O$_{5+z}$**  P4/mmm  (space group 123)
a = 3.87655 Å   c = 7.67781 Å   V = 115.379 Å$^3$
Z = 1   chi$^2$ = 7.32

| Atom | Wyckoff position | x | y | z | Å$^2$ |
|------|------------------|-----|-----|-------|------|
| Pr | 1a | 0 | 0 | 0 | 1.21 |
| Ba | 1b | ½ | ½ | ½ | 1.21 |
| Co | 2h | ½ | ½ | 0.259 | 1.34 |
| O | 1c | ½ | ½ | 0 | 1.75 |
| O | 1d | ½ | ½ | ½ | 1.75 |
| O | 4i | ½ | 0 | 0.232 | 1.75 |

dCo-O : 1.8504 Å (1 coordinate)   1.9494 Å (4 coordinate)   1.9886 Å (1 coordinate)
dPr – O : 2.6325 Å (8 coordinate)   2.7412 Å (4 coordinate)
dBa – O : 2.8269 Å (8 coordinate)   2.7412 Å (4 coordinate)
No rotation of octahedrons (forbidden by symmetry), thus, 180°

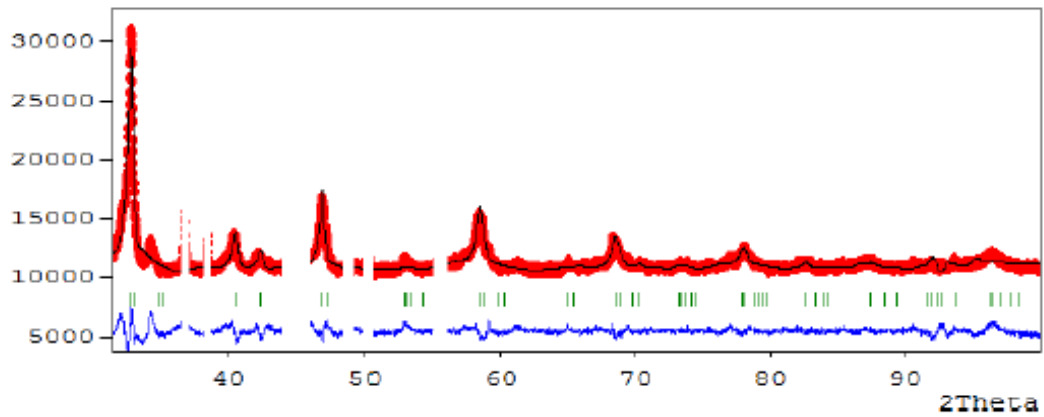

## S5. Electrochemical Analysis

*Tafel Plot – Chronoamperometry Study*

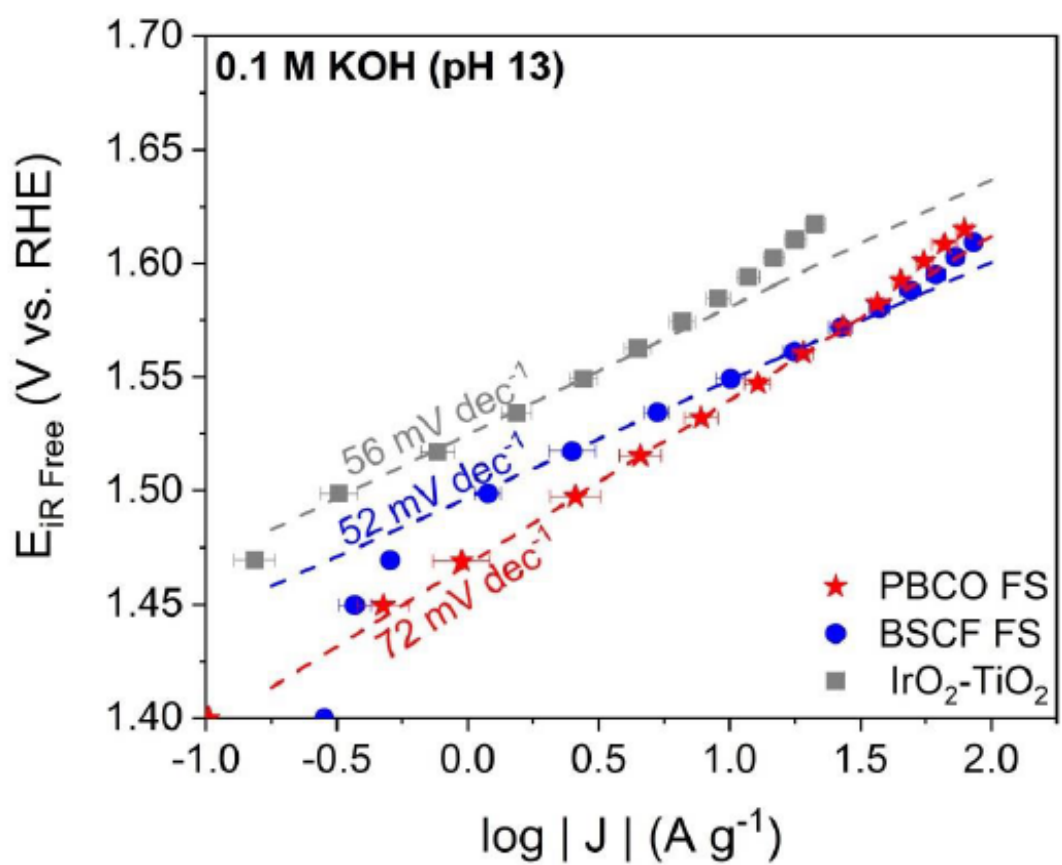

**Figure S7.** Tafel plots comparing OER activities of BSCF-FS, PBCO-FS and commercial $IrO_2$.

In **Figure S7**, both BSCF-FS and PBCO-FS show lower overpotentials than that of the $IrO_2$-$TiO_2$. This trend of activity is also verified from the galvodynamic test of the alkaline exchange membrane water electrolyzer (AEMWE) (refer to Figure 5).

## S6. Galvanodynamic Test in Alkaline Membrane Water Electrolyzer

Referring to the RDE study (Figure S3), BSCF-FS reveals increasing current density over cyclic voltammetry cycles. This increase of activity is also observed when BSCF-FS is adopted as the anodic electrode in a real alkaline water electrolyzer (see **Figure S8**). In case of PBCO-FS, the improvement of activity in AMWE is not as significant as that of BSCF-FS. Note that Figure 4 of the main text compares polarization curves obtained after 1 minute.

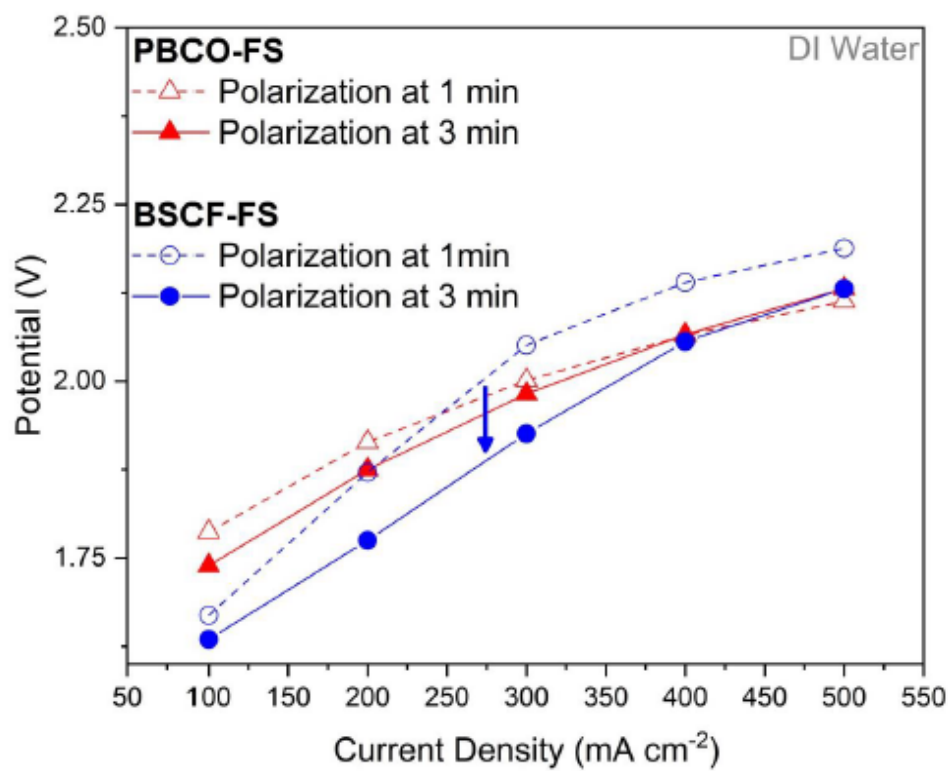

**Figure S8.** Comparison of polarization curves obtained after 1 minute and 3 minutes for MEAs having BSCF-FS and PBCO-FS as anodic electrodes.